\documentclass[11pt]{article}
\usepackage{amsfonts}
\usepackage{lscape}
\usepackage{latexsym,amsmath}
\usepackage{amssymb,array}
\usepackage{subfigure}
\makeatletter 
\RequirePackage[dvips]{graphicx} 
\newtheorem{t1}{Theorem}[section]

\begin{document}
\title{\textbf{Some estimators of the PMF and CDF of the Logarithmic Series Distribution}}
\author{Sudhansu S. Maiti\footnote{Corresponding author. e-mail:
dssm1@rediffmail.com}, Indrani Mukherjee and Monojit Das\\
Department of Statistics, Visva-Bharati
University, Santiniketan-731 235, West Bengal, India}
\date{}
\maketitle
%\newpage
%\begin{center}
%\textbf{Acknowledgement}

%\vspace{0.2in}

%I would like to acknowledge my respected teachers for providing me enough guidance to complete my project work easily and also I would like to give thanks to all of my seniors and friends who helps me cordially to complete my task.
%\end{center}
%\newpage
\begin{center}
Abstract
\end{center}
This article addresses the different methods of estimation of the probability mass function (PMF) and the cumulative distribution function (CDF) for the Logarithmic Series distribution. Following estimation methods are considered: uniformly minimum variance unbiased estimator (UMVUE), maximum likelihood estimator (MLE), percentile estimator (PCE), least square estimator (LSE), weighted least square estimator (WLSE). Monte Carlo simulations are performed to compare the performances of the proposed methods of estimation. 
%A real data set has been analysed for illustrative purpose.

\vspace{0.5cm}

\noindent \textbf{Keywords:}  Maximum likelihood estimators; uniformly minimum variance unbiased estimators; least square estimators; weighted least square estimators; percentile estimators.
\\ {\bf 2010 Mathematics Subject Classification.} 62E15, 62N05. 
\newpage
\section{Introduction}
\ A random variable $X$ is said to have the Logarithmic series distribution, if its probability mass function (PMF) is given by
\begin{eqnarray}\label{logpmf}
P(X=x;p)=f(x)=\frac{-1}{\ln (1-p)}\frac{p^x}{x},~ x=1,2,\ldots; 0<p<1
\end{eqnarray}
and its cumulative distribution function (CDF) is given by
\begin{eqnarray}
F(x)=\sum_{w=1}^x \frac{-1}{\ln (1-p)}\frac{p^w}{w},~ x=1,2,\ldots; 0<p<1.
\end{eqnarray}
\ The above distribution has many application in biology and ecology. It is also used for modelling data linked to the number of species in occupancy-abundance studies. It is a limiting case of the zero truncated negative binomial distribution. The univariate Logarithmic series distribution was brought to light by Fisher [\ref{fishar}]. He was then dealing with the biological data on species and individual collected by Corbet and Williams [\ref{fishar}]. The Logarithmic series distribution found its way further in meteorology [Williams (\ref{willi1952}), Ramabhadran (\ref{ram})], in human ecology [Clark, Eckstron and Linden (\ref{clark})] and in operation research [Williamson and Bretherton (\ref{bretherton}), Haight and Reichenbach (\ref{haight})]. Patil, Kamat and Wani(\ref{patil}) have studied extensively the structure and statistics of the logarithmic series distribution. Nelson and David (\ref{nelson}) also have a technical report to their credit on the subject.

\par Now a days researchers have given attention for study of properties and inference on this distribution. Some extension models have been found out and their properties and statistical inferences are made by host of authors. Statisticians are most of the times interested about inferring the parameter(s) involved in the distribution. MLE and Bayes estimate of the parameter has been focused by the authors. Hardly any unbiased estimator of the parameter has been studied so far and finding out minimum variance unbiased estimator (MVUE) of the parameter seems to be intractable and consequently the comparison with any unbiased class of estimator is not being made. However instead of studying the estimators of the parameters, we have scope to find out unbiased estimator of the PMF and the CDF as well as some biased estimator of the same is possible and comparison among the estimators could be made. This is why we have shifted our focus from estimation of parameter to estimation of the PMF and the CDF.
\par We see many situations where we have to estimate PMF, CDF or both. For instance, PMF can be used for estimation of differential entropy, R\'{e}nyi entropy, Kullback-Leibler divergence and Fisher information; CDF can be used for estimation of cumulative residual entropy, the quantile function, Bonferroni curve, Lorenz curve and both PMF and CDF can be used for estimation of probability weighted moments, hazard rate function, mean deviation about mean etc.
\par Some studies on the estimation of PDF and CDF have appeared in recent literature for some continuous distributions: Pareto distribution [Asrabadi (\ref{Asrabadi-1990}), Dixit and Jabbari (\ref{Dixit-Nooghabi-2010}), Dixit and Jabbari (\ref{Dixit-Jabbari-2011})], Exponential Pareto distribution [Jabbari and Jabbari (\ref{Jabbari- Jabbari-2010})], Generalized Rayleigh distribution [Alizadeh (\ref{ali})], Weibull extension model [Bagheri et al. (\ref{weibull})], Exponentiated Gumbel distribution [Bagheri et al. (\ref{Bagheri-Alizadeh-Nadarajah-2016})], Generalized Exponential-Poisson distribution [Bagheri et al. (\ref{Bagheri-Alizadeh-Jamkhaneh-Nadarajah-2013})], Generalized Exponential distribution [Alizadeh et al.(\ref{gen})], Lindley distribution [Maiti and Mukherjee (\ref{indu1})]. Work towards discrete random variable like Binomial, Poisson, Geometric, Negative Binomial, the estimation issues of the PMF and CDF have been made by Maiti et al. (\ref{maiti}).
\par The work has been organised as following. In section 2, Maximum likelihood estimation of the PMF and CDF have been found out substituting the MLE of the parameter by the virtue of Invariance property. Section 3 covered all the calculations about Uniformly Minimum Variance Unbiased Estimator of the PMF and the CDF and the theoretical expressions of variance of these estimators. Least squares, weighted least squares and percentile estimators have included in sections 4 and 5. After that section 6 covered simulation study result and at the end we give the conclusion in section 7.
\section{Maximum likelihood estimators of the PMF and the CDF}
\ Let $X_1, X_2,...,X_n$ be a random sample of size n with PMF $\eqref{logpmf}$. The MLE of the above distribution is being deducted.
\begin{eqnarray*}
L(\theta;x)&=&\left(\frac{-1}{\ln (1-p)}\right)^n p^{\sum_{i=1}^n x_i}\frac{1}{\prod_{i=1}^n x_i}\\
\ln L(\theta;x)&=& -n\ln(-\ln (1-p))+\sum_{i=1}^n x_i \ln p-\sum_{i=1}^n \ln x_i\\
\end{eqnarray*}
\begin{eqnarray}
\frac{dl(\theta)}{d\theta}&=&0\nonumber\\
\Rightarrow \frac{n}{(1-p)\ln (1-p)}&=&-\frac{1}{p}\sum_{i=1}^n x_i\nonumber\\
\Rightarrow \frac{1}{\ln (1-p)^{\frac{1-p}{p}}}&=&-\frac{T}{n}\nonumber\\
\Rightarrow (1-p)^{\frac{1-p}{p}}&=& e^{-\frac{n}{T}}
\end{eqnarray}
Here we can not find out any simple expression for MLE of $p$.
Therefore, by using numerical approach we have to find out the root of the equation. 

\section{\textbf{UMVU estimators of the PMF and the CDF}}
\ In this section, we obtain the UMVU estimators of the PMF and the CDF of the Logarithmic Series distribution. Also, we obtain the MSEs of these estimators.
\par Here $T=\sum_{i=1}^n X_i$ is a complete sufficient statistic for $p$. Being a sum of n independent random variables with logarithmic Series distribution with the parameter $p$ has Stirling distribution of the first kind $SDFK(n,p)$ [Johnson et al.($\ref{Johnson2005}$)], $T$ has the following mass function 
\begin{equation}\label{gt}
P(T_X=x; n, p)= \frac{n!|s(x,n)|p^x}{x!(-\ln (1-p))^n},~x=n,n+1,\ldots
\end{equation}
\par Let $X_1, X_2,..., X_n$ be a random sample of size $n$ from Logarithmic Series distribution given by equation ($\ref{logpmf}$). 
Then, $T$ is a complete sufficient statistic for $p$ and its PMF is given by ($\ref{gt}$). According to Rao-Blackwell theorem and Lehmann-Scheffe theorem, we get the UMVUE of the PMF and the CDF.
\par Consider,
\begin{eqnarray*}
Y&=&1~ \mbox{if}~X_i=k\\
&=&0~ \mbox{otherwise}
\end{eqnarray*}
Then 
\begin{equation*}
E_p(Y)=P_p[X_1=k]=\gamma (p) ~ \mbox{for all}~p.
\end{equation*}
where $\gamma (p)=\frac{-1}{\ln (1-p)}\frac{p^k}{k}$.
Define, 
\begin{eqnarray}\label{umvue}
E\left[Y|t\right]&=&P_p\left[X_1=k|T=t\right]\nonumber \\
&=&\frac{P_p\left[X_1=k,~ \sum_{i=2}^n X_i=t-k\right]}{P_p\left[T=t\right]}\nonumber \\
&=&\frac{1}{nk}\times \frac{|s(t-k,n-1)|}{|s(t,n)|}\times \frac{t!}{(t-k)!};~k=n,n+1,\ldots , t
\end{eqnarray}
Here $E\left[Y|t\right]$ is the UMVUE of the PMF of the aforesaid distribution.

\vspace{0.2in}

\begin{t1}\label{umvuefandF}
Let $T=t$ be given. Then
\begin{equation*}
\widehat{f}(x)=\frac{1}{nx}\times \frac{|s(t-x,n-1)|}{|s(t,n)|}\times \frac{t!}{(t-x)!};~x=n,n+1,\ldots , t
\end{equation*}
is the UMVUE for $f(x)$ and
\begin{equation*}
\widehat{F}(x)=\sum_{w=n}^x \frac{1}{nw}\times \frac{|s(t-w,n-1)|}{|s(t,n)|}\times \frac{t!}{(t-w)!};~x=n,n+1,\ldots , t
\end{equation*}
is the UMVUE for $F(x)$.
\end{t1}

\vspace{0.2in}

\textbf{Proof:}

\vspace{0.2in}

\ The proof of $\widehat{f}(x)$ is the UMVUE follows from equation($\ref{umvue}$). Similarly the proof that $\widehat{F}(x$) is the UMVUE of $\widehat{F}(x)$.

\vspace{0.2in}

\begin{t1}

\vspace{0.2in}

The variance of $\widehat{f}(x)$ is given by
\begin{equation*}
Var(\widehat{f}(x))=\frac{(n-1)!}{x^2[-\ln (1-p)]^n}\left[\frac{1}{n} \sum_{t=x}^\infty \frac{|s(t-x,n-1)|^2}{|s(t,n)|}\times  \frac{t!p^t}{((t-x)!)^2}-\frac{(n-1)!}{(-\ln (1-p))^n}\left\lbrace \sum_{t=x}^\infty \frac{p^t |s(t-x,n-1)|}{(t-x)!}\right\rbrace^2\right]
\end{equation*}
\ and
\begin{eqnarray*}
Var(\widehat{F}(x))&=&\frac{(n-1)!}{(-\ln (1-p))^n}[\frac{1}{n}\sum_{t=x}^\infty \frac{t!p^t}{|s(t,n)|}\left\lbrace \sum_{w=n}^x\frac{|s(t-w,n-1)|}{w(t-w)!}\right\rbrace^2 -\frac{(n-1)!}{(-\ln (1-p))^n}\\
&&\times \left\lbrace \sum_{t=x}^\infty \sum_{w=n}^x\frac{p^t|s(t-w,n-1)|}{w(t-w)!}\right\rbrace^2]
\end{eqnarray*}
\end{t1}

\vspace{0.2in}

\textbf{Proof:}

\vspace{0.2in}

Using the Theorem $\ref{umvuefandF}$. We can get the value of $Var(\widehat{f}(x))$. The expression for the variance for 
$\widehat{f}(x)$ follows by $Var(\widehat{f}(x))=E(\widehat{f}(x))^2-E^2(\widehat{f}(x))$.

\vspace{0.2in}

Therefore,
\begin{eqnarray*}
Var(\widehat{f}(x))&=&E(\widehat{f}(x))^2-E^2(\widehat{f}(x))\\
&=&\sum_{t=x}^\infty (\widehat{f}(x))^2g_T(t)-\left[\sum_{t=x}^\infty \widehat{f}(x)g_T(t)\right]^2\\
&=&\left[\sum_{t=x}^\infty \frac{1}{n^2x^2}\times \frac{|s(t-x,n-1)|^2}{|s(t,n)|^2}\times \frac{t!^2}{(t-x)!^2}\times \frac{n!|s(t,n)|p^t}{t!(-\ln (1-p))^n}\right]\\
&&-\left[ \sum_{t=x}^\infty \frac{1}{nx}\times \frac{|s(t-x,n-1)|}{|s(t,n)|}\times \frac{t!}{(t-x)!}\times \frac{n!|s(t,n)|p^t}{t!(-\ln (1-p))^n}\right]^2\\
&=&\frac{(n-1)!}{x^2[-\ln (1-p)]^n}[\frac{1}{n} \sum_{t=x}^\infty \frac{|s(t-x,n-1)|^2}{|s(t,n)|}\times  \frac{t!p^t}{((t-x)!)^2}\\
&&-\frac{(n-1)!}{(-\ln (1-p))^n}\left\lbrace \sum_{t=x}^\infty \frac{p^t |s(t-x,n-1)|}{(t-x)!}\right\rbrace^2]
\end{eqnarray*}
\ Where,~ $g_T(t)=\frac{n!|s(t,n)|p^t}{t!(-\ln (1-p))^n}$. The proof for the expression of variance for $\widehat{F}(x)$ is similar.
%\newpage
%\begin{figure}[hbtp]
%\centering
%\includegraphics[scale=1]{includevarpmftheo.pdf}
%\caption{Graph of variance of UMVU estimator of PMF}
%\end{figure}
%\begin{figure}[hbtp]
%\centering
%\includegraphics[scale=1]{varCDFtheo.pdf}
%\caption{Graph of UMVU estimator of CDF}
%\end{figure}

\begin{figure}[ht]
\centering
\subfigure{\includegraphics[height=2in,width=2.5in]{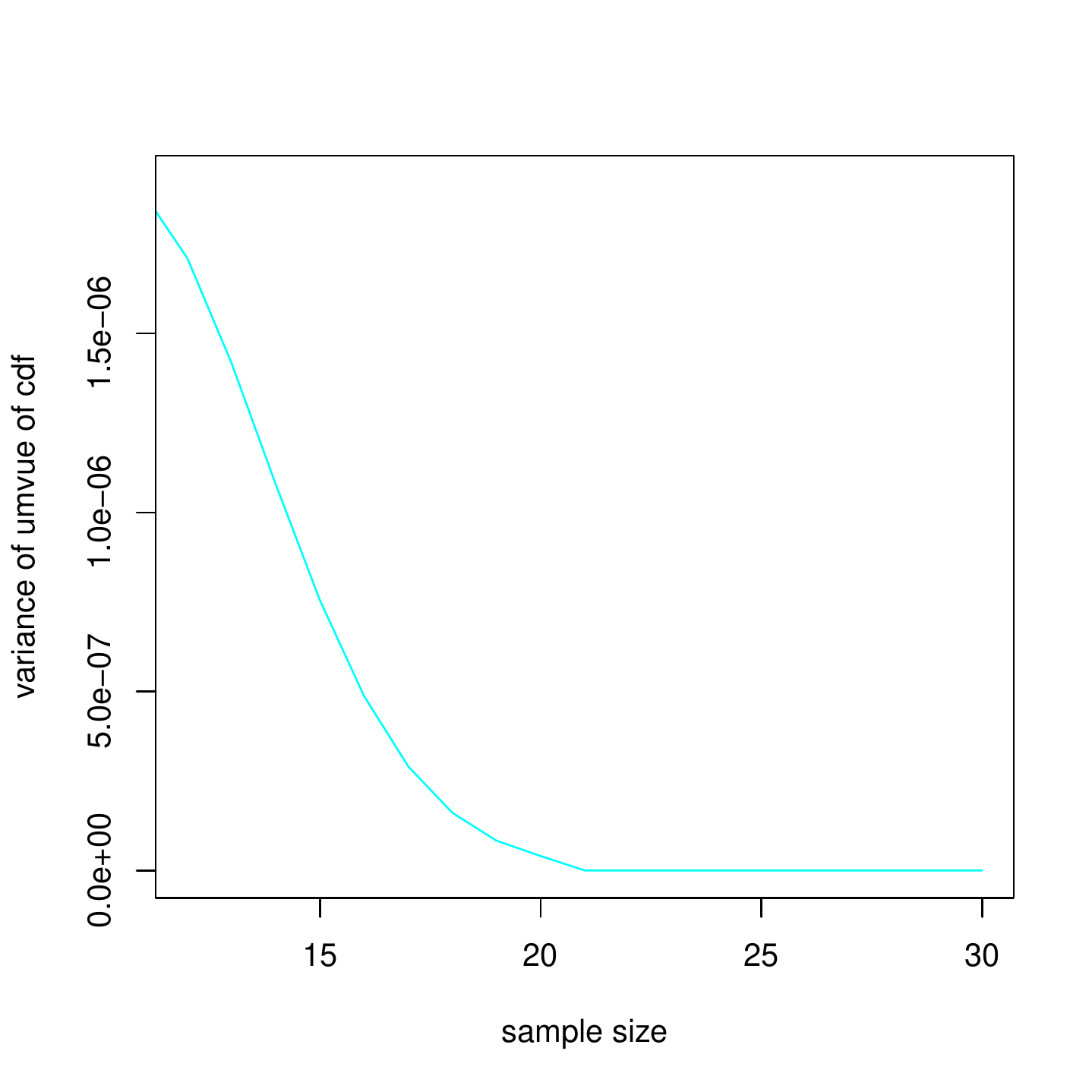}}
\subfigure{\includegraphics[height=2in,width=2.5in]{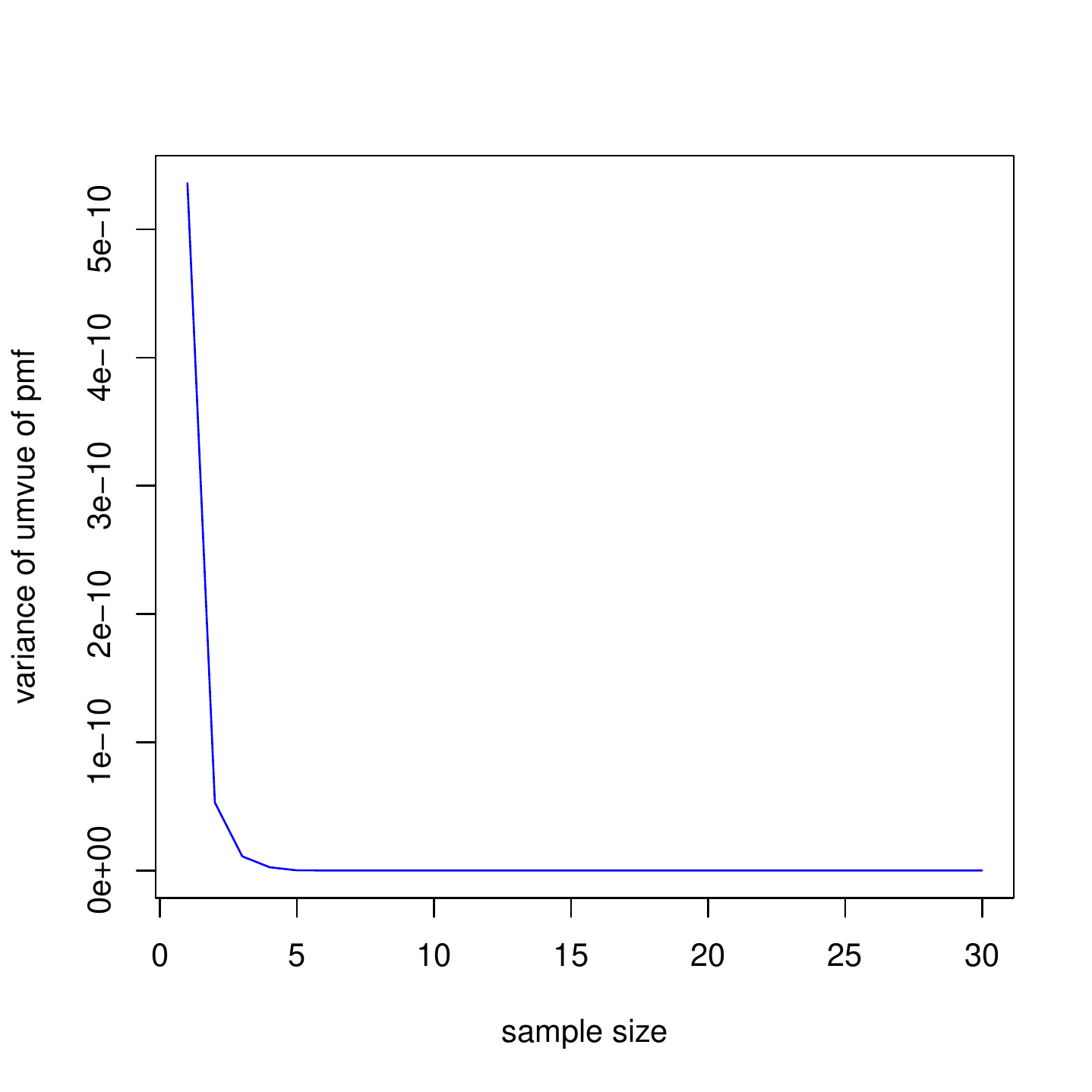}}
\caption[]{Graph of variance of UMVU estimator of the CDF and the PMF.}
\end{figure}

%\newpage
\section{\textbf{Least squares and weighted least squares estimators}}
\ The least square estimators and weighted least square estimators were proposed by Swain et al. [\ref{Sw88}] to estimate the parameters of Beta distributions. In this paper, we apply the same technique for the Logarithmic Series distribution. Suppose $X_1,..., X_n$ is a random sample of size $n$ from a CDF $F(.)$ and let $X_{i:n}$, $i=1,...,n$ denote the ordered sample in ascending order. The proposed method uses the CDF of $F(x_{i:n})$. For a sample of size $n$, we have $E[F(X_{j:n})]=\frac{j}{n+1}$, $Var[F(X_{j:n})]=\frac{j(n-j+1)}{(n+1)^2(n+2)}$ and $Cov[F(X_{j:n}),F(X_{k:n})]=\frac{j(n-k+1)}{(n+1)^2(n+2)}$ for $j<k$, (see Johnson et al. [$\ref{Johnson-Kotz-Balakrishnan-1994}$]). Using the expectations and the variances, two variants of the least squares method follow.

\vspace{0.2in}

\textbf{Method 1: Least squares estimators}

\vspace{0.2in}

This method is based on minimizing   

\vspace{0.2in}

$\sum_{j=1}^n [F(x_{j:n})-\frac{j}{n+1}]^2$

\vspace{0.2in}

with respect to the unknown parameters. 

\vspace{0.2in}

\ In case of Logarithmic Series Distribution the least squares estimators of p is $\widetilde{p}_{LSE}$.
$\widetilde{p}_{LSE}$ can be obtained by minimizing

\vspace{0.2in}

$\sum_{j=1}^n[\sum_{w=1}^{x(j)} \frac{-1}{\ln (1-p)}\frac{p^w}{w}-\frac{j}{n+1}]^2$
with respect to $\theta$.

\vspace{0.2in}

So, to obtain the  LS estimators of the PMF and the CDF, we use the same method as for the MLE. Therefore,
\begin{equation}
\widetilde{f}_{LSE}(x)=\frac{-1}{\ln (1-\widetilde{p}_{LSE})}\frac{\widetilde{p}_{LSE}^x}{x}
\end{equation}
and
\begin{equation}
\widetilde{F}_{LSE}(x)=\sum_{w=1}^x \frac{-1}{\ln (1-\widetilde{p}_{LSE})}\frac{\widetilde{p}_{LSE}^w}{w}
\end{equation}
It is difficult to find the expectations and the MSE of these estimators analytically, so we calculate them by means of simulation study.

\vspace{0.2in}

\textbf{Method 2: Weighted Least squares estimators}

\vspace{0.2in}

This method is based on minimizing

\vspace{0.2in}

$\sum_{j=1}^n w_j[F(x_{j:n})-\frac{j}{n+1}]^2$

\vspace{0.2in}

with respect to the unknown parameters, where

\vspace{0.2in}

$ w_j=\frac{1}{Var[F(X_{j:n})]}=\frac{(n+1)^2(n+2)}{j(n-j+1)}$

\vspace{0.2in}

\ In case of the Logarithmic Series distribution, the weighted least squares estimators of $p$ say 
$\widetilde{p}_{WLSE}$ is the value minimizing 

\vspace{0.2in}

$\sum_{j=1}^n w_j[\sum_{w=1}^{x(j)} \frac{-1}{\ln (1-p)}\frac{p^w}{w}-\frac{j}{n+1}]^2$.

\vspace{0.2in}

So, the WLS estimators of the PMF and CDF are
\begin{equation}
\widetilde{f}_{WLSE}(x)=\frac{-1}{\ln (1-\widetilde{p}_{WLSE})}\frac{\widetilde{p}_{WLSE}^x}{x}
\end{equation}
and
\begin{equation}
\widetilde{F}_{WLSE}(x)=\sum_{w=1}^x \frac{-1}{\ln (1-\widetilde{p}_{WLSE})}\frac{\widetilde{p}_{WLSE}^w}{w}
\end{equation}
It is difficult to find the expectations and the MSE of these estimators analytically. So, we can calculate them by means of a simulation study. 

\section{\textbf{Estimators based on percentiles}}
\ Estimations based on percentiles was originally  suggested by Kao [$\ref{Ka58}$, $\ref{Ka59}$].
Percentiles estimators are based on inverting the CDF. Since the Logarithmic Series distribution has a closed form CDF, its parameters can be estimated using percentiles.
This method is based on minimizing
\begin{equation*}
\sum_{i=1}^n [\ln p_i - \ln F(x_{(i)})]^2
\end{equation*}
Let $X_{i:n}$, $i=1,...,n$ denote the ordered random sample from the Logarithmic Series distribution. Also let $p_i=\frac{i}{n+1}$.
The percentile estimator of $p$ say $\widetilde{p}_{PCE}$ is the value minimizing

\vspace{0.2in}

$\sum_{i=1}^n [ \ln p_i-\ln \left\lbrace\sum_{w=1}^{x(j)} \frac{-1}{\ln (1-p)}\frac{p^w}{w}\right\rbrace]^2$.

\vspace{0.2in}

So, the percentile estimators of the PMF and CDF are
\begin{equation}
\widetilde{f}_{PCE}(x)=\frac{-1}{\ln (1-\widetilde{p}_{PCE})}\frac{\widetilde{p}_{PCE}^x}{x}
\end{equation}
and
\begin{equation}
\widetilde{F}_{PCE}(x)=\sum_{w=1}^x \frac{-1}{\ln (1-\widetilde{p}_{PCE})}\frac{\widetilde{p}_{PCE}^w}{w}
\end{equation}
The expectations and the MSE of these estimators can be calculated by simulation.

\section{Simulation study}
\ Here, we conduct Monte Carlo simulation to evaluate the performance of the estimators for the PMF and the CDF discussed in the previous sections. All computations were performed using the R-software. We evaluate the performance of the estimators based on MSEs. The MSEs were computed by generating $1000$ replications, taking $p=0.6, x=12$, from Logarithmic series Distribution. It is observed that MSEs decreases with increasing sample size. It verifies the consistency properties of all the estimators. We observe from true MSE point of view, UMVUE is better than MLE for both PMF and CDF. We have generated observations using the algorithm given by Kemp[$\ref{Kemp}$].
%\begin{figure}[hbtp]
%\centering
%\includegraphics[scale=1]{msefx.pdf}
%\caption{MSEs of MLE, UMVUE, LSE, WLSE, PCE for the PMF and the CDF for the Logarithmic Series Distribution}
%\end{figure}
%\begin{figure}[hbtp]
%\centering
%\includegraphics[scale=1]{mseFFx.pdf}
%\caption{MSEs of MLE, UMVUE, LSE, WLSE, PCE for the PMF and the CDF for the Logarithmic Series Distribution}
%\end{figure}

\begin{figure}[ht]
\centering
\subfigure{\includegraphics[height=2in,width=2.5in]{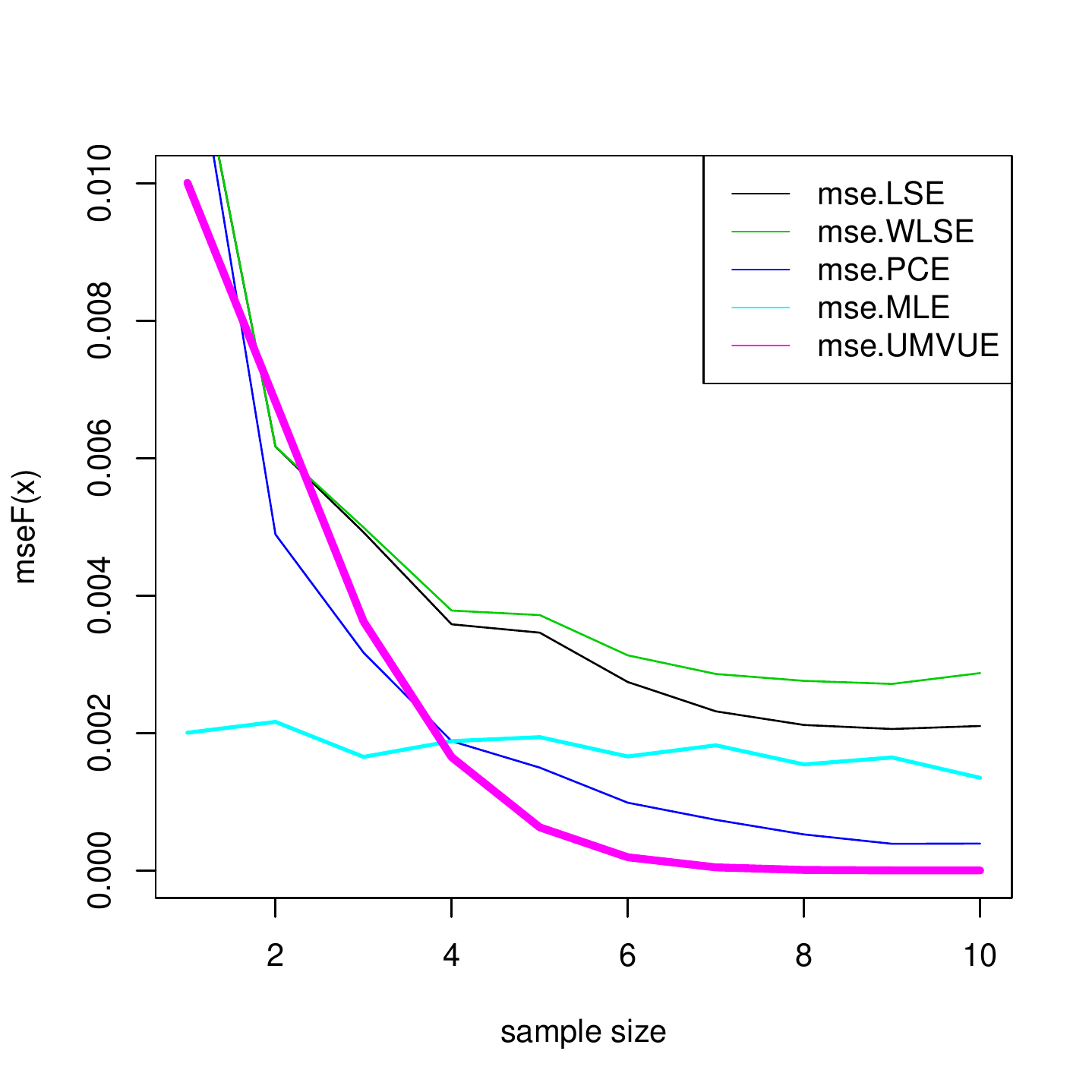}}
\subfigure{\includegraphics[height=2in,width=2.5in]{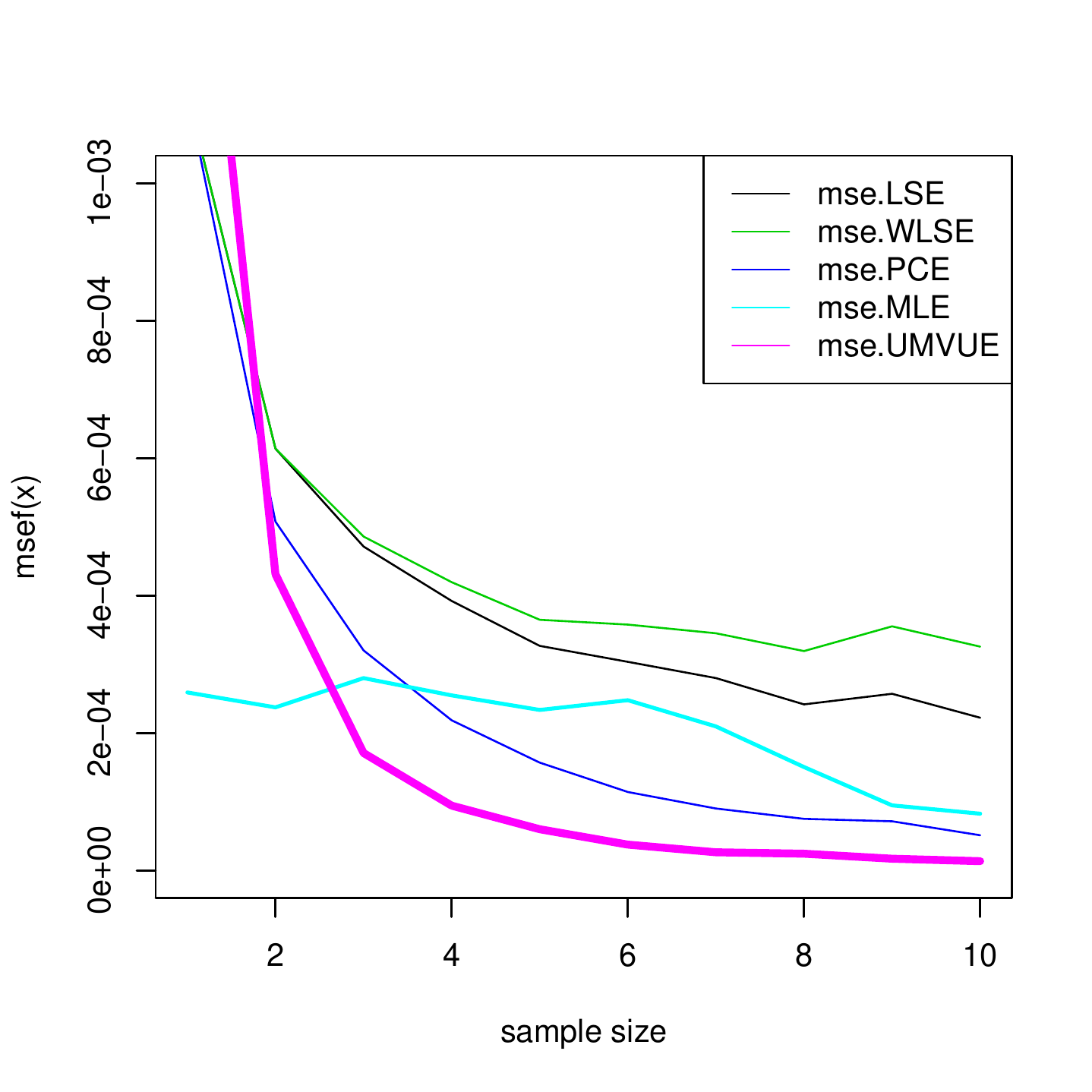}}
\caption[]{MSEs of MLE, UMVUE, LSE, WLSE, PCE for the CDF and the PMF for the Logarithmic Series Distribution.}
\end{figure}

%\section{Application}

\newpage
\section{Conclusion}
\ Here different methods of estimation of the PMF and the CDF of the Logarithmic Series distribution have been considered. Uniformly minimum variance unbiased estimator (UMVUE), maximum likelihood estimator (MLE), percentile estimator (PCE), least square estimator (LSE) and weighted least square estimator (WLSE) have been found out. Monte Carlo simulations are performed to compare the performances of the proposed methods of estimation. If we restrict to unbiased class of estimators, UMVUE is better
in minimum variance sense. For small samples MLE is better though it is biased.

\end{document}